\documentclass{article}
\usepackage{amsmath,amsfonts}
\usepackage{algorithmic}
\usepackage{algorithm}
\usepackage{array}
\usepackage{textcomp}
\usepackage{stfloats}
\usepackage{xcolor}
\usepackage{url}
\usepackage{verbatim}
\usepackage{graphicx}
\usepackage[skip=0.333\baselineskip]{caption}
\usepackage{subcaption}
\usepackage{booktabs}
\usepackage{multirow}
\usepackage[affil-it]{authblk}
\usepackage{IEEEtrantools}

\IEEEeqnarraydefcolsep{0}{\hoffset}

\usepackage[square,numbers]{natbib}
\newcommand{\NaturalNumbers}{\mathbb{N}}

\newcommand{\Time}{t}
\newcommand{\TimeSet}{\mathcal{T}}

\newcommand{\NodeSet}{\mathcal{V}}
\newcommand{\EdgeSet}{\mathcal{E}}

\newcommand{\ExternalNode}{i}
\newcommand{\ExternalNodeSet}{\mathcal{I}}

\newcommand{\InternalNode}{p}
\newcommand{\InternalNodeAlt}{q}
\newcommand{\InternalNodeSet}{\mathcal{P}}

\newcommand{\ConverterEdge}{\ExternalNode\InternalNode}
\newcommand{\ConverterEdgeSet}{\mathcal{E}^{\text{i}}}

\newcommand{\BreakerEdge}{\InternalNode\InternalNodeAlt}
\newcommand{\BreakerEdgeAlt}{\InternalNodeAlt\InternalNode}
\newcommand{\BreakerEdgeSet}{\mathcal{E}^{\text{p}}}

\newcommand{\Contingency}{f}
\newcommand{\ContingencySet}{\mathcal{F}}

\newcommand{\Converter}{\mathsf{x}}
\newcommand{\Breaker}{\mathsf{y}}
\newcommand{\Power}{\mathsf{P}}
\newcommand{\PowerImbalance}{\Power^{\text{imb.}}}
\newcommand{\PowerImbalancePositive}{\Power^{\text{imb.+}}}
\newcommand{\PowerImbalanceNegative}{\Power^{\text{imb.-}}}

\newcommand{\NetPosition}{P^{\text{net pos.}}}
\newcommand{\MinimumPower}{P^{\text{min}}}
\newcommand{\MaximumPower}{P^{\text{max}}}

\newcommand{\NumberOfBreakers}{N^{\text{cb}}}
\newcommand{\Cost}{c}
\newcommand{\BreakerCost}{\Cost^{\text{CB}}}

\newcommand{\ImbalanceCostPositive}{\Cost^{\text{imb.+}}}
\newcommand{\ImbalanceCostNegative}{\Cost^{\text{imb.-}}}
\newcommand{\ReferenceValue}{\Cost^{\text{ref.}}}

\newcommand{\Probability}{\varphi}

\newcommand{\MTTF}{\text{MTTF}}
\newcommand{\Length}{L}
\newcommand{\Duration}{d}

\newcommand{\MW}{\text{ MW}}
\newcommand{\km}{\text{ km}}

\newcommand{\riska}{1752}
\newcommand{\riskb}{886}
\newcommand{\riskc}{800}
\newcommand{\riskd}{723}
\newcommand{\riske}{697}
\newcommand{\risklowprobnocb}{525.6\: \ReferenceValue}
\newcommand{\risklowprobonecb}{265.8 \: \ReferenceValue}

\newcommand{\cbone}{886}
\newcommand{\cbtwo}{86}
\newcommand{\cbthree}{77}
\newcommand{\cbfour}{26}

\begin{document}

\title{Topology Optimization for DC Circuit Breaker Placement in HVDC Switching Stations}

\author[1,2]{Merijn Van Deyck}
\author[3]{Tom Van Acker}
\author[1,2]{Geraint Chaffey}
\author[1,2]{Dirk Van Hertem}

\affil[1]{Department of Electrical Engineering, KU Leuven, Belgium}
\affil[2]{{Etch-EnergyVille, Belgium}}
\affil[3]{{Elia Transmission, Belgium}}

\date{}
\maketitle

\begin{abstract}
HVDC protection will be required in future multiterminal HVDC grids to prevent large outages caused by DC faults. Therefore, system-level protection design is essential for the development of HVDC switching stations that connect several converter stations and lines within these grids. This paper presents an optimization method for the design of HVDC circuit breaker (DCCB) configurations in HVDC switching stations and electrical energy hubs. This approach builds on the current practice of using selected configurations based on pre-defined protection strategies. In contrast to these existing methods, the DC switching station design in the proposed method offers significantly more flexibility and allows the consideration of large numbers of relevant operating conditions, leading to more effective, optimal design outcomes. A mixed-integer linear optimization problem is formulated to design the DC protection and minimize the risk of high impact DC faults. An example case study demonstrates that the optimization method allows the calculation of the optimal number of DCCBs for a given DC switching station, based on the failure rates of DC grid components and the DCCB cost relative to the fault impact. With these results, the marginal benefit to risk reduction of each additional DCCB included in a DC switching station is calculated. Moreover, the result of the optimization problem provides the \textit{optimal} breaker configuration for the required number of DCCBs and can consequently be used as a topological design tool for DC switching stations.
\end{abstract}
\section{Introduction}
\label{Introduction}

High voltage DC (HVDC) grids play a crucial role in the transition towards a decarbonized energy system. Their ability to transmit power over large distances is a key benefit for the integration of remote renewable energy sources and for grid reinforcement and interconnections~\cite{MianReview}. 
Recent plans for new HVDC projects have increasingly shifted towards modular multilevel converter-based multiterminal HVDC grids that bring together several DC connections in \textit{electrical energy hubs} (EEH)~\cite{Definitions} or DC switching stations (DCSS)~\cite{InterOpera3_1}. This shift from point-to-point links to multiterminal connections increases flexibility, efficiency and robustness~\cite{DirkObstacles}. Moreover, an integrated offshore multiterminal HVDC grid consisting of several offshore EEHs has been identified as a preferred solution for the integration and further development of offshore wind farms in the North Sea~\cite{TYNDP_North_Sea}.
\subsection{HVDC Protection}
While the development of HVDC grids connected through DCSS is expected to improve socio-economic welfare and facilitate the integration of renewable energy sources~\cite{TYNDP_2024}, a key technical -- and economic -- challenge remains in the risk for high-impact contingencies in DC grids. This risk, which has been a key obstacle for the development of multiterminal HVDC grids for many years~\cite{DirkObstacles}, stems from the nature of fault currents caused by DC-side short-circuit faults. Such fault currents are, due to rapid capacitive discharge of cables and converters in the DC grid and the lack of natural zero-crossings, impossible to interrupt with ``traditional'' switchgear as used in AC grid protection systems~\cite{leterme2019designing}.
Therefore, a non-selective approach to HVDC protection has been applied in nearly all existing HVDC systems with only one exception currently operating in China~\cite{Jovcic2019AdoptingGrids,Zhangbei2021}.

Non-selective protection systems typically rely on the interruption of fault currents at the AC-side of converters using AC circuit breakers. During this process, the converters bypass their internal power electronic components and capacitors to avoid damage, known as  \textit{blocking}. The converters then act as uncontrolled rectifiers into the DC grid~\cite{leterme2016equivalent}. During this period, the DC grid voltage cannot be maintained and there can be no controlled active power flow through the HVDC grid. Thus, the entire HVDC grid is temporarily unable to operate until the faulted element is cleared from the de-energized DC grid, after which the remaining healthy components can be restored to normal operation.

This non-selective approach to HVDC protection has proved sufficient for point-to-point, cabled HVDC systems. In these systems, any DC-side fault leads to the permanent outage of the entire link (or the faulted pole in case of pole-to-ground faults in bipolar systems with metallic return), and the speed and location of the fault interruption are of little consequence. However, the loss of an entire multiterminal HVDC grid, even for a short time, is likely to have an intolerably high impact on the frequency and voltage angle stability of connected AC grids~\cite{DCcont2017KUL}. Therefore, protection strategies that rely on HVDC circuit breakers (DCCB) to interrupt fault currents within the DC grid have been proposed. DCCBs facilitate selective fault clearing within the HVDC grid, allowing healthy DC grid elements to remain in continuous operation~\cite{leterme2015classification}. The need for such selective strategies has been identified for HVDC grids that consist of multiple offshore hubs~\cite{SuperGridInstitute} and in general for multiterminal HVDC grids~\cite{Dave2022Thesis}. However, although DCCBs are approaching technological maturity and global commercial availability~\cite{DCCB_TRL}, the cost of these components is expected to remain a significant barrier to their widespread implementation.

\begin{figure}[h]
    \centering
    \includegraphics[width=0.8\linewidth]{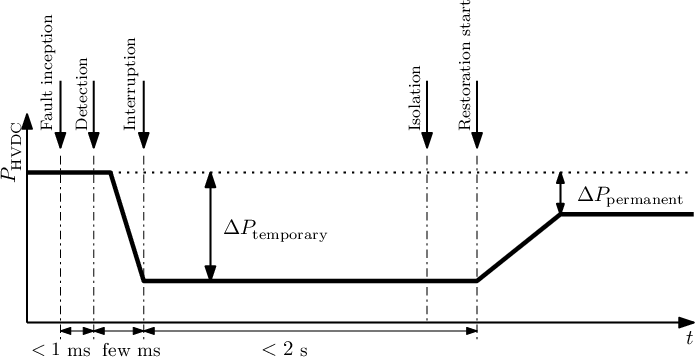}
    \caption{Example representation of the total power transfer through an HVDC grid during the DC fault clearing sequence, assuming partially selective protection and automatic reconnection of converters after temporary stops (timings are indicative).}
    \label{fig:protection_sequence}
\end{figure}

To mitigate the high investment costs required for a fully selective protection system that protects each element individually using DCCBs, a \textit{partially selective} strategy can instead be applied. This strategy relies on the use of fewer DCCBs, not to rapidly clear any faulted element from the rest of the system, but to split the grid into several protection zones, keeping the non-faulted zones in continuous operation and de-energizing and restoring the faulted zone. This temporary de-energization causes a temporary loss of power transfer capacity during the time between the fault interruption and the restoration of the non-faulted elements, after galvanic isolation of the fault at zero current. Following the DC grid restoration at the end of the fault clearing process, a permanent loss of power transfer capacity remains due to the isolation of the faulted element from the grid. In case of permanent DC faults, such as isolation breakdown in a cable, the permanent loss of power transfer capacity persists until the faulted element is repaired. Therefore, the selectivity of the DC protection system, defined by the placement of DCCBs, mainly impacts the magnitude of the temporary loss of power transfer during the fault clearing process, as illustrated in Fig.~\ref{fig:protection_sequence}.

\subsection{Problem Statement}
\begin{figure}[h]
    \centering
    \begin{subfigure}{0.32\linewidth}
        \includegraphics[width=\linewidth]{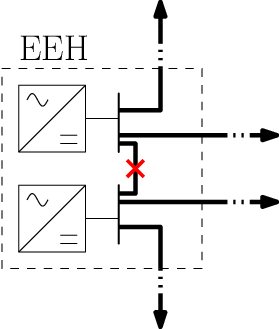} 
        \caption{}
        \label{fig:PS1}
    \end{subfigure}\hfill
    \begin{subfigure}{0.32\linewidth}
    \includegraphics[width=\linewidth]{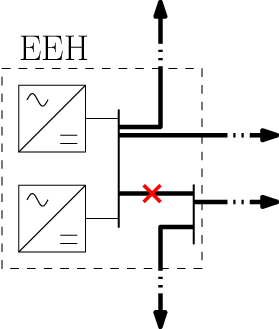}
    \caption{}
    \label{fig:PS2}
    \end{subfigure}\hfill
    \begin{subfigure}{0.32\linewidth}
    \includegraphics[width=\linewidth]{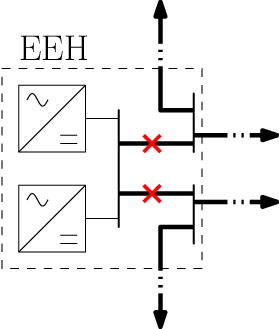}
    \caption{}
    \label{fig:PS3}
    \end{subfigure}
    \caption{Different possible `partially selective' breaker configurations in an example EEH}
    \label{fig:DCCB_config_examples}
\end{figure}

The choice of a protection strategy is a fundamental aspect of multiterminal HVDC grid design. Several methods have been proposed to consider the trade-off between \textit{investment cost} and the \textit{AC grid frequency stability in case of faults}~\cite{MerijnTrondheim}. However, in the context of HVDC grids that consist of several links connected to a central substation - a DCSS or EEH - simply determining a protection strategy does not fully determine the design. For example, a large DCSS could be ``split'' in many ways when applying the partially selective strategy, using different numbers of DCCBs in various configurations, as illustrated in Fig.~\ref{fig:DCCB_config_examples}. In fact, even if the number of DCCBs is determined in the design, thus fixing the expected investment costs, many possible configurations of the substation can still be achieved by arranging the DCCBs and busbar connections in different ways. Even with the same number of DCCBs, these different configurations could offer different levels of fault-impact-mitigation, depending on the power flows and anticipated faults in the HVDC grid~\cite{MerijnShortlisting}. Therefore, this paper proposes an optimization method for the design of DCSS topologies, focusing on the placement - or \textit{configuration} - of DCCBs, aiming to minimize the risk of temporary power imbalances caused by DC grid contingencies in all relevant operational scenarios. 

\subsection{Literature Review}
The novelty of the proposed method, compared to existing optimization-based grid design methods, lies in the consideration of fault impacts caused by the transient power imbalance in the objective, while remaining fully flexible in the DCSS topology and protection system design. Typical optimization tools that consider the effect of contingencies on the grid operation, so-called Security-Constrained Optimal Power Flow (SCOPF) methods, focus on the steady-state operating point of the grid before and after relevant contingencies~\cite{capitanescu2011state}. Although additional time-domain studies may be performed to validate the frequency stability of considered systems after a fault, or constraints ensuring stability can be included in these methods, they inherently focus on optimizing the operating point of the grid, e.g. by minimizing generation costs or losses, and are usually less focused on minimizing the impact of faults through optimal protection system design~\cite{saplamidis2015security}. Security constraints can also be included in the grid design stage, rather than the operational stage, in Transmission Network Expansion Planning (TNEP) problems, as proposed in~\cite{dominguez2017milp} for hybrid AC and DC systems. Similarly to typical SCOPF methods, this optimization method focuses on steady-state aspects, such as the availability of connections after contingencies and the need for redundancy in the grid planning, rather than the design of protection systems, which do not influence the long-term availability of connections but impact the grid stability on a transient time scale. 

Several studies have specifically considered the design of HVDC protection systems in an Optimal Power Flow (OPF) framework, evaluating the benefits of using DCCBs in multiterminal HVDC grids. In~\cite{maciver2015reliability}, the reliability, costs and losses of several HVDC grids with different protection system implementations are compared in order to determine the economic benefit of using DCCBs. This benefit is represented by the fact that DCCBs allow a more interconnected HVDC grid without risking the maximum Loss of Infeed (LoI) limit being exceeded in connected AC grids. Each case considered in this study is designed such that LoI limits in connected AC grids are always respected. In this regard, fault impacts caused by transient power imbalances are again represented only as a constraint while the benefit of DCCBs is considered in the possibility to connect more lines together while staying within this security limit. Similarly,~\cite{MerijnACDC2025} applies OPF calculations to compare a multiterminal HVDC grid using one DCCB to a similar system that is preventively decoupled when LoI constraints could be exceeded, thus evaluating the economic benefit of a selective protection system without considering the actual impact of transient power imbalances caused by DC faults.

A more detailed representation of fault impacts on the AC grid frequency stability is considered in the TNEP method presented in~\cite{Dave2022Thesis}. A frequency response model is used to represent the impact of DC faults under different protection strategies. The impact of DC faults is represented economically by the cost of reserves needed to mitigate AC grid imbalances and avoid under-frequency load shedding. In this way, the fault impact is compared to the expected investment costs of DCCBs. By representing the fault impact as a cost, the performance of the DC protection system can be considered in the objective of the optimization. However, the protection system itself is not fully flexible for design in~\cite{Dave2022Thesis}, as only a fully selective and non-selective strategy are considered, without flexibility for the placement and number of DCCBs.

\subsection{Contribution and Outline}
As the AC grid inertia provided by conventional generators continues to decrease worldwide, designing protection systems based on existing LoI constraints, which ensure frequency stability can be maintained today, may not prove sufficient for the AC grid of the future. In this future system, transient power imbalances that are deemed acceptable in current grid codes may cause much more severe frequency instabilities due to the reduced amount of inertia. Therefore, when designing protection systems for multiterminal HVDC grids, optimizing the DCCB configuration to \textit{minimize} possible fault impacts ensures the effectiveness of the protection system. Moreover, if the impact of transient power imbalances is represented by a cost, a cost-benefit analysis can be performed to determine the optimal number of DCCBs, which impose significant investment costs, resulting in optimal protection system configurations for HVDC switching stations. 

Therefore, this paper proposes a Mixed-Integer Linear Programming (MILP) optimization framework for the design of the topology and DCCB placement in HVDC switching stations, such that power imbalances caused by DC grid faults are minimized. The principles of the proposed method build on those used in~\cite{MerijnShortlisting}, where an iterative, graph model-based method was used to shortlist DCCB configurations for specific EEH case studies. The optimization method discussed in this paper formulates the principles governing the operation of HVDC protection systems in a structured mathematical framework. A risk-based design approach is presented and illustrated for a case study. This approach integrates aspects like the DC cable failure rate, DCCB costs and realistic power flow setpoints in the design of the DC protection system. 
Consequently, the proposed optimization method forms the basis for a flexible tool for protection system design in HVDC switching stations, which can be adapted based on the needs of system developers and used for the cost-efficient development of secure HVDC grids in the future.

As indicated by the literature review, the combined focus on transient fault impacts in the objective and the fully flexible design of DC protection configurations differentiates this method from existing grid design methods. The proposed method also improves on existing expert-based design methods for HVDC protection systems which focus on selecting and implementing predefined protection strategies. In this paper, a more flexible approach that leads to an optimal design is taken. 

The conceptual principles and mathematical framework of the optimization problem are explained in Section~\ref{sec:mathematical_framework}. Section~\ref{sec:case_study_introduction} then introduces a case study for the application of the method, followed by the discussion of the results in Section~\ref{sec:results}. Section~\ref{sec:conclusion} then draws conclusions and discusses future work.

\section{Mathematical Framework}
\label{sec:mathematical_framework}
This section presents the mathematical framework for the DCCB placement optimization problem. This framework must accurately represent the operation of the DC grid and the protection system, before and during contingencies, as well the fault impact and the design constraints used to generate optimal DCSS topologies in the solution. Tab.~\ref{tab:notation} provides a general overview and generic examples of the notation used throughout this section to describe the formulation.
\begin{table}[ht!]
    \centering
    \caption{Notation used for mathematical framework}
    \label{tab:notation}
    \begin{tabular}{ l l c }
        element         & notation                  & example                           \\ \hline
        index           & lower-case italic         & $x$                               \\
        set             & calligraphic              & $\mathcal{X}$                     \\ 
        variable        & sans serif                & $\mathsf{X}$                      \\
        parameter       & italic                    & $X$
    \end{tabular}
\end{table}

\subsection{Sets, Variables and Parameters}
\label{sec:mathematical_notation}

\begin{figure}[b!]
    \centering
    \includegraphics[scale=1]{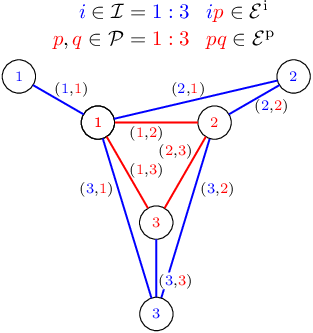}
    \caption{Illustration of the graph representing a DC switching station.}
    \label{fig:dc_substation}
\end{figure}

A DCSS is represented in the optimization problem by a graph~${(\NodeSet,\EdgeSet)}$ with vertices~${\NodeSet=\ExternalNodeSet \cup \InternalNodeSet}$ and edges~${\EdgeSet=\ConverterEdgeSet\cup\BreakerEdgeSet}$ (Fig.~\ref{fig:dc_substation}). Nodes are grouped into two unique subsets: external AC nodes~${\ExternalNode \in \ExternalNodeSet \subset \NaturalNumbers}$, which are connected to the DCSS through a converter station and HVDC line, and internal DC nodes within the switching station~${\InternalNode \in \InternalNodeSet \subset \NaturalNumbers}$. Similarly, the edges are also grouped into two unique subsets: connections~${\ConverterEdge \in \ConverterEdgeSet}$, i.e. DC cables and converters, between an external node~$\ExternalNode$ and a DC node~$\InternalNode$, and DC breakers~${\BreakerEdge \in \BreakerEdgeSet}$ between two internal DC nodes~$\InternalNode$ and~$\InternalNodeAlt$. As illustrated in Fig.~\ref{fig:dc_substation}, the connection set $\ConverterEdgeSet$ contains all edges $\ConverterEdge$ for which ${\InternalNode \leq \ExternalNode}$, which limits the number of binary variables and symmetrical solutions to the optimization problem. This reduction of the problem size stems from the assumption that all internal nodes ${\InternalNode \in \InternalNodeSet}$ can be considered as identical DC connection points in the DCSS. Moreover, as DCCBs are assumed to work bidirectionally, the set of breaker edges $\BreakerEdgeSet$ contains only edges $\BreakerEdge$ for which ${\InternalNode < \InternalNodeAlt}$.

In addition to the node and edge sets, two other sets are defined: discrete points in time are denoted by~${\Time \in \TimeSet}$ and DC faults are denoted by~${\Contingency \in \ContingencySet = \ExternalNodeSet \subset \NaturalNumbers}$.

A binary variable~$\Converter_{\ConverterEdge}$ is introduced for each possible connection~${\ConverterEdge \in \ConverterEdgeSet}$ to indicate whether it exists (${\Converter_{\ConverterEdge}\!=\!1}$) or not (${\Converter_{\ConverterEdge}\!=\!0}$). The power flow through a connection~$\ConverterEdge\! \in\! \ConverterEdgeSet$ from external node~$\ExternalNode$ to internal node~$\InternalNode$ at a time~$\Time\! \in\! \TimeSet$ is denoted as~$\Power_{\ConverterEdge,\Time}$, and as~$\Power_{\ConverterEdge,\Time,\Contingency}$ for a given contingency~${\Contingency\! \in\! \ContingencySet}$. Similarly, for each possible DC breaker~$\BreakerEdge\! \in\! \BreakerEdgeSet$, a binary variable~$\Breaker_{\BreakerEdge}$ and power flow variables~$\Power_{\BreakerEdge,\Time}$ and~$\Power_{\BreakerEdge,\Time,\Contingency}$ are defined. The power flow through a breaker or connection is limited between~$\MinimumPower$ and~$\MaximumPower$. Lastly, the net position of an external node~${\ExternalNode\! \in\! \ExternalNodeSet}$ at a discrete time~${\Time\! \in\! \TimeSet}$, i.e. how much power is demanded from or supplied to the DC grid at this node, is denoted as~$\NetPosition_{\ExternalNode,\Time}$. Furthermore, for every AC node~${\ExternalNode\! \in\! \ExternalNodeSet}$, a power imbalance variable~$\PowerImbalance_{\ExternalNode,\Time,\Contingency}$ is defined for every discrete time~$\Time\! \in\! \TimeSet$ and contingency~$\Contingency\! \in\! \ContingencySet$. This variable can subsequently be split into positive and negative parts:~$\PowerImbalancePositive_{\ExternalNode,\Time,\Contingency}$ and~$\PowerImbalanceNegative_{\ExternalNode,\Time,\Contingency}$, respectively. Note that at all times, at least either the positive or negative part should be zero.  

\subsection{Constraints}
\label{sec:constraints}
The constraints to the optimization problem, listed in~\eqref{eq:single_converter_edge}-\eqref{eq:DCCB_faulted_power_flow}, determine the design rules for the optimal DCCB configuration, while also representing the power flows and the operation of the DC protection system. It is determined by~\eqref{eq:single_converter_edge} that only one connection may exist between each external node and the DCSS. Subsequent constraints~\eqref{eq:normal_operation_power_balance}-\eqref{eq:power_imbalance} enforce the power balance at the external AC nodes during normal and contingent operation, matching the flow through the HVDC grid (and any power imbalances) to the given net position in each time step. Similarly, \eqref{eq:internal_normal_power_balance} and \eqref{eq:internal_contingent_power_balance} enforce the nodal power balance for the internal DC nodes. 

Connection constraints~\eqref{eq:con_limit_normal} and~\eqref{eq:con_limit_contingent} pose boundaries on the power flow through connections $\ConverterEdge \in \ConverterEdgeSet$. In~\eqref{eq:con_limit_protection}, the operation of the protection system is represented. This constraint enforces that, if a faulted connection $\Contingency\! \in\! \ContingencySet$ is connected to a certain DC node $\InternalNode \in \InternalNodeSet$ (i.e. $\Converter_{\Contingency\InternalNode} = 1$), no power may flow, during contingent operation, through any connection $\ConverterEdge$ connected to this DC node $\InternalNode$. Consequently,~\eqref{eq:con_limit_protection} represents the operation of a DC protection system that de-energizes the entire faulted protection zone, represented here by a DC node, thus interrupting power flow through all connections to that node.

The maximum number of DCCBs in the switching station is constrained by~\eqref{eq:DCCB_count}. The constraints in~\eqref{eq:DCCB_power_normal} and~\eqref{eq:DCCB_power_cont} limit the power flow through breaker edges in normal and contingent operation respectively. Finally,~\eqref{eq:DCCB_faulted_power_flow} represents the opening of DCCBs, allowing no power flow through breaker $\BreakerEdge$ if either $\InternalNode$ or $\InternalNodeAlt$ (denoted in~\eqref{eq:DCCB_faulted_power_flow} with the auxiliary index $k \in \{\InternalNode,\InternalNodeAlt\}$) is connected to the faulted line. 

\begin{IEEEeqnarray}{ 0 l }
    \textsc{External Node Constraints} \nonumber \\
    \text{converter per ext. node limit --~} \forall \ExternalNode \in \ExternalNodeSet: \nonumber\\
    \quad \sum_{\InternalNode \in \InternalNodeSet\: \text{if}\: \InternalNode \leq \ExternalNode} \Converter_{\ConverterEdge} = 1, \label{eq:single_converter_edge} \\
    \text{normal operation power balance --~} \forall \ExternalNode \in \ExternalNodeSet, \Time \in \TimeSet: \nonumber \\
    \quad \sum_{\InternalNode \in \InternalNodeSet\: \text{if}\: \InternalNode \leq \ExternalNode} \Power_{\ConverterEdge,\Time} = \NetPosition_{\ExternalNode,\Time} \label{eq:normal_operation_power_balance} \\
    \text{contingent power balance --~} \forall \ExternalNode \in \ExternalNodeSet, \Time \in \TimeSet, \Contingency \in \ContingencySet: \nonumber \\
    \quad \sum_{\InternalNode \in \InternalNodeSet\: \text{if}\: \InternalNode \leq \ExternalNode} \Power_{\ConverterEdge,\Time,\Contingency} = \NetPosition_{\ExternalNode,\Time} + \PowerImbalance_{\ExternalNode,\Time,\Contingency} \label{eq:contingency_power_balance} \\
    \text{power imbalance --~} \forall \ExternalNode \in \ExternalNodeSet, \Time \in \TimeSet, \Contingency \in \ContingencySet: \nonumber \\
    \quad \PowerImbalance_{\ExternalNode,\Time,\Contingency} = \PowerImbalancePositive_{\ExternalNode,\Time,\Contingency} + \PowerImbalanceNegative_{\ExternalNode,\Time,\Contingency} \label{eq:power_imbalance} \\
    \textsc{Internal Node Constraints} \nonumber \\
    \text{normal operation power balance --~} \forall \InternalNode \in \InternalNodeSet, \Time \in \TimeSet: \nonumber \\
    \quad \sum_{\ExternalNode \in \ExternalNodeSet} \Power_{\ConverterEdge,\Time} = \sum_{\InternalNodeAlt \in \InternalNodeSet \: \text{if}\: \InternalNodeAlt > \InternalNode} \Power_{\BreakerEdge,\Time} - \sum_{\InternalNodeAlt \in \InternalNodeSet\:\text{if}\:\InternalNodeAlt<\InternalNode} \Power_{\BreakerEdgeAlt,\Time} \label{eq:internal_normal_power_balance}\\
    \text{contingent power balance --~} \forall \InternalNode \in \InternalNodeSet, \Time \in \TimeSet, \Contingency \in \ContingencySet: \nonumber \\
    \quad \sum_{\ExternalNode \in \ExternalNodeSet} \Power_{\ConverterEdge,\Time,\Contingency} = \sum_{\InternalNodeAlt \in \InternalNodeSet \: \text{if}\: \InternalNodeAlt > \InternalNode} \Power_{\BreakerEdge,\Time,\Contingency} - \sum_{\InternalNodeAlt \in \InternalNodeSet\:\text{if}\:\InternalNodeAlt<\InternalNode} \Power_{\BreakerEdgeAlt,\Time,\Contingency} \label{eq:internal_contingent_power_balance} \\
    \textsc{Connection Constraints} \nonumber \\
    \text{connection power limit --~} \forall \ConverterEdge \in \ConverterEdgeSet, \Time \in \TimeSet: \nonumber \\
    \Converter_{\ConverterEdge}\,\MinimumPower_{\ConverterEdge} \leq \Power_{\ConverterEdge,\Time} \leq \Converter_{\ConverterEdge}\,\MaximumPower_{\ConverterEdge} \label{eq:con_limit_normal}\\
    \text{~~~~~~~~~~~~~~~~~~~~~~~~~~~ --~} \forall \ConverterEdge \in \ConverterEdgeSet, \Time \in \TimeSet, \Contingency \in \ContingencySet: \nonumber \\
    \Converter_{\ConverterEdge}\,\MinimumPower_{\ConverterEdge} \leq \Power_{\ConverterEdge,\Time,\Contingency} \leq \Converter_{\ConverterEdge}\,\MaximumPower_{\ConverterEdge} \label{eq:con_limit_contingent}\\
    \text{~~~~~~~~~ --~} \forall \ExternalNode \in \ExternalNodeSet, \InternalNode \in \InternalNodeSet, \Time \in \TimeSet, \Contingency \in \ContingencySet~\text{if}~\InternalNode \leq \ExternalNode~\text{and}~\InternalNode \leq \Contingency: \nonumber \\
    (1-\Converter_{\Contingency\InternalNode})\,\MinimumPower_{\ConverterEdge} \leq \Power_{\ConverterEdge,\Time,\Contingency} \leq (1-\Converter_{\Contingency\InternalNode})\,\MaximumPower_{\ConverterEdge} \label{eq:con_limit_protection}\\
    \textsc{DCCB Constraints} \nonumber \\
    \text{DCCB count limit}: \nonumber \\
    \sum_{\BreakerEdge \in \BreakerEdgeSet} \Breaker_{\BreakerEdge} \leq \NumberOfBreakers \label{eq:DCCB_count} \\
    \text{DCCB power limit --~} \forall \BreakerEdge \in \BreakerEdgeSet, \Time \in \TimeSet: \nonumber \\
    \Breaker_{\BreakerEdge}\,\MinimumPower_{\BreakerEdge} \leq \Power_{\BreakerEdge,\Time} \leq \Breaker_{\BreakerEdge}\,\MaximumPower_{\BreakerEdge} \label{eq:DCCB_power_normal}\\
    \text{~~~~~~~~~~~~~~~~~~~~~~~ --~} \forall \BreakerEdge \in \BreakerEdgeSet, \Time \in \TimeSet, \Contingency \in \ContingencySet: \nonumber \\
    \Breaker_{\BreakerEdge}\,\MinimumPower_{\BreakerEdge} \leq \Power_{\BreakerEdge,\Time,\Contingency} \leq \Breaker_{\BreakerEdge}\,\MaximumPower_{\BreakerEdge} \label{eq:DCCB_power_cont} \\
    \text{~~~~~~~~~~~~ --~}\forall \BreakerEdge \in \BreakerEdgeSet, k \in \{\InternalNode,\InternalNodeAlt\}, \Time \in \TimeSet, \Contingency \in \ContingencySet~\text{if}~k \leq \Contingency: \nonumber \\
    (1-\Converter_{\Contingency k})\MinimumPower_{\BreakerEdge} \leq \Power_{\BreakerEdge,\Time,\Contingency} \leq (1-\Converter_{\Contingency k})\MaximumPower_{\BreakerEdge} 
    \label{eq:DCCB_faulted_power_flow}
\end{IEEEeqnarray}

\subsection{Objective function}
\label{sec:objective}
Given the high costs of DCCBs, the optimization problem can be structured as a cost-benefit analysis to determine the optimal number of breakers to include -- in an optimal configuration -- in the DCSS. To this end, the total DCCB cost is compared to the \textit{risk} related to power imbalances caused by DC faults. In~\eqref{eq:CBA_objective}, it is shown how this trade-off between investment cost and risk is represented in the objective function of the optimization problem. The first term in~\eqref{eq:CBA_objective} represents the total capital costs for the time period $\TimeSet$, assuming a fixed cost $\BreakerCost$ per DCCB. The second term in~\eqref{eq:CBA_objective} represents the total risk. The impact of DC faults is quantified using a cost for the positive and negative power imbalance ($\ImbalanceCostPositive$ and $\ImbalanceCostNegative$) for each external AC node. For every fault $\Contingency \in \ContingencySet$, this cost-based representation of the impact is multiplied with the probability of this fault occurring ($\Probability_\Contingency$) during a given time step $\Time\! \in\! \TimeSet$ to represent the total risk. In this paper, the probability of each fault is assumed constant in time, though variable failure rates could also be considered. 
\begin{IEEEeqnarray}{ 0 l }
    \min \nonumber \\
    \sum_{\BreakerEdge \in \BreakerEdgeSet} \BreakerCost\Breaker_{\BreakerEdge} +    \sum_{\ExternalNode \in \ExternalNodeSet, \Time \in \TimeSet,\Contingency \in \ContingencySet}\Probability_{\Contingency} \left(\ImbalanceCostPositive_{\ExternalNode,\Time}\PowerImbalancePositive_{\ExternalNode,\Time, \Contingency}  - \ImbalanceCostNegative_{\ExternalNode,\Time}\PowerImbalanceNegative_{\ExternalNode,\Time,\Contingency} \right) \label{eq:CBA_objective}
\end{IEEEeqnarray}
\section{Case Study Description}
\label{sec:case_study_introduction}
\begin{figure}[ht!]
    \centering
    \includegraphics[scale=1]{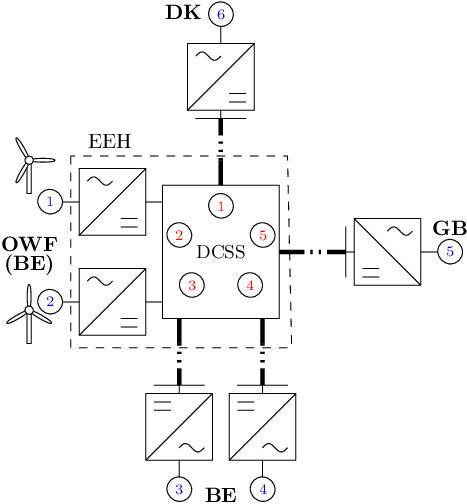}
    \caption{Schematic representation of the case study network. Internal 
    DC nodes $\InternalNodeSet$ are shown in red and external nodes $\ExternalNodeSet$ are shown in blue.}
    \label{fig:case_study_network}
\end{figure}

This section presents the network and input data for the illustrative case study used to demonstrate the DCCB configuration optimization. A schematic representation of the considered network is given in Fig.~\ref{fig:case_study_network}. In the following subsections, the test case is described conceptually and linked to the parameters and sets used in the optimization problem.
\vspace{-0.5cm}
\subsection{HVDC grid layout}
The presented case study considers an HVDC grid, containing a DCSS connected to six external AC nodes: ${\ExternalNodeSet\! =\! 1\!:\!6}$. The DCSS is assumed to be located offshore, in close proximity to two offshore wind farms, with identical generation profiles, represented by two external nodes: ${\ExternalNode\! =\! 1}$ and ${\ExternalNode\! =\! 2}$. The other external nodes represent onshore AC grids.  

Every external AC node is connected to the DCSS by a bipolar HVDC converter station and cable, representing the connections ${\ConverterEdge\! \in \!\ConverterEdgeSet}$. These bipolar connections each have a total capacity of ${2000\MW}$. However, as only pole-to-ground cable faults are considered in the case study, a single pole representation of the grid is used in the optimization problem. Therefore, the minimum and maximum power ratings can be expressed as ${\MinimumPower_{\ConverterEdge} = -1000\MW}$ and ${\MaximumPower_{\ConverterEdge} = 1000\MW}$, for all ${\ConverterEdge \in \ConverterEdgeSet}$. Each external AC node is located at a certain distance from the DCSS. Tab.~\ref{tab:lengths} shows the lengths of connections between each external node and the DCSS.

\begin{table}[ht!]
    \centering
    \caption{Lengths of connections from external nodes $\ExternalNode \in \ExternalNodeSet$ to the DCSS}
    \label{tab:lengths}
    \begin{tabular}{ c c c}
        External AC node         & Length $\Length_\ExternalNode$ [km]   \\ \hline
        $\ExternalNode = 1$           & 0                               \\
        $\ExternalNode = 2$           & 0                      \\ 
        $\ExternalNode = 3$           & 70                         \\ 
        $\ExternalNode = 4$           & 100                        \\ 
        $\ExternalNode = 5$           & 150                         \\ 
        $\ExternalNode = 6$           & 700                         \\ 
    \end{tabular}
\end{table}

\subsection{Contingencies}
The optimization problem considers a set of contingencies $\ContingencySet$ that, in this case study, represent pole-to-ground cable faults in the HVDC grid. The fault probability $\Probability_\Contingency$ for each contingency can be derived from Mean Time To Failure (MTTF) data from literature~\cite{maciver2015reliability,Li2015Reliability,Wang2016Reliability}. The MTTF, typically expressed for HVDC cables in years/$100\km$, can be converted to the probability of a fault $\Contingency$ occurring in any time step $\Time$ through~\eqref{eq:MTTF_to_prob}, where $\Length_\Contingency$ is the length of the faulted connection (in km) and $\Duration_\Time$ the duration of any time step ${\Time \in \TimeSet}$ (in hours).
\begin{equation}
    \label{eq:MTTF_to_prob}
    \Probability_\Contingency = \frac{1}{\MTTF}\cdot \frac{\Length_\Contingency}{100}\cdot \frac{\Duration_\Time}{8760}
\end{equation}
For the example case study in Section~\ref{sec:CBA_case_study}, an example MTTF of 15~years/$100\km$ is used~\cite{Wang2016Reliability}. Using a time step duration of $\Duration_\Time = 1\text{ hour}$, this leads to a fault probability per time step of ${\Probability_\Contingency = 7.61\cdot 10^{-8}\cdot \Length_\Contingency}$.

Since the probability of DC cable faults depends on the connection length, and the DCSS is, in this case study, assumed to be close to the offshore wind farms represented by external nodes ${\ExternalNode \!=\! 1}$ and ${\ExternalNode\! =\! 2}$, the optimization problem can be simplified by not considering any faults on these connection. Consequently, instead of ${\ContingencySet\! =\! \ExternalNodeSet}$, the set of contingencies becomes ${\ContingencySet = \{3:6\} \subset \ExternalNodeSet}$.

\subsection{Pre-fault power flows}
The power flow variables in the optimization problem depend on the net positions in the external AC nodes, given by $\NetPosition_{\ExternalNode,\Time}$. These must be provided as input to the optimization problem for each node $\ExternalNode \in \ExternalNodeSet$ and time step $\Time\! \in\! \TimeSet$. To gather this input data for the case study, a simple economic dispatch optimization was run using publicly available data and projections of load and renewable energy source profiles, generation costs and installed capacities~\cite{TYNDP_2024}. By including a variable that represents power import and export for each market zone in the dispatch model, the optimal power transfer through the HVDC grid is calculated for each time step $\Time \in \TimeSet$. Moreover, as the dispatch problem relies on data for European countries and market zones~\cite{TYNDP_2024}, the external AC nodes $\ExternalNode \in \ExternalNodeSet$ must be matched to the considered countries to match  $\NetPosition_{\ExternalNode,\Time}$ to the results of the dispatch optimization. In the case study, the offshore wind farms represented by nodes $\ExternalNode\! =\! 1$ and $\ExternalNode\! =\! 2$ are located near the Belgian coast. External nodes $\ExternalNode\! =\! 3$ and $\ExternalNode\! =\! 4$ represent different locations in the Belgian onshore grid, with identical demand profiles. Two other external grids are considered: $\ExternalNode\! =\! 5$ represents the GB power system and $\ExternalNode\! =\! 6$ the continental part of the Danish power system.

\subsection{Fault impact cost representation}
To accurately represent the impact of transient power imbalances caused by DC faults -- and affected by the DCCB configuration -- the positive and negative imbalance costs $\ImbalanceCostPositive$ and $\ImbalanceCostNegative$ must quantify the monetary impact caused by the temporary loss of power transfer. This transient power imbalance could cause a large rate of change of frequency in AC grids, possibly leading to frequency instability. However, reserves markets meant to prevent such instabilities (FCR and aFRR procurement markets) are not fully suitable to represent the imbalance costs in this paper. These markets are independent of the \textit{occurrence} of a fault, and instead procure some amount of capacity for every market interval~\cite{EntsoE_FCR}. Alternatively, costs related to the \textit{activation} of reserves (e.g. aFRR energy costs) do depend on the occurrence of imbalances in the grid~\cite{Elia_aFRR}. However, the impact of the transient power imbalance caused by DC faults extends beyond the energy cost (which is expected to be relatively small due to the short duration of this stage in the overall fault clearing process, as illustrated in Fig.~\ref{fig:protection_sequence}), as even short imbalances can affect frequency stability, depending on the magnitude of the imbalance and the AC grid inertia~\cite{DCcont2017KUL}. As a result, a detailed cost representation that accurately quantifies the impact of the occurrence of a transient power imbalance should be developed in future studies in order to use the proposed optimization method in a comprehensive cost-benefit analysis for the optimization of the DCCB count and configuration in a real system. However, since this is out of the scope of this paper, example values for $\ImbalanceCostPositive$ and $\ImbalanceCostNegative$ are used for the illustration of the method. Tab.~\ref{tab:imbalance_costs} shows how the positive and negative imbalance costs in each considered external node relate to the reference value $\ReferenceValue$. Although it is possible to consider time-dependent imbalance costs, constant values are used here. In general, it is assumed for this case study that negative imbalance costs are lower than positive imbalance costs. The positive imbalance cost for $\ExternalNode\! =\! 1$ and $\ExternalNode\! =\! 2$ is assumed relatively high since these external nodes represent offshore wind farms that cannot increase their power output in case of a fault. For external nodes ${\ExternalNode\!=\!3}$ and ${\ExternalNode\!=\!4}$, imbalance costs are assumed identical as these nodes are located near each other. 

\begin{table}[ht!]
    \centering
    \caption{Relative imbalance costs for each external zone}
    \label{tab:imbalance_costs}
    \begin{tabular}{ c c c}
        External AC node         & $\ImbalanceCostPositive_{\ExternalNode}$ [$\ReferenceValue$] & $\ImbalanceCostNegative_{\ExternalNode}$ [$\ReferenceValue$]\\ \hline
        $\ExternalNode = 1$           & $100 $&$ 0.2 $                              \\
        $\ExternalNode = 2$           & $100 $& $0.2 $                   \\ 
        $\ExternalNode = 3$           & $1$ & $0.5$                        \\ 
        $\ExternalNode = 4$           & $1$ & $0.5$                              \\ 
        $\ExternalNode = 5$           & $1.5$ & $0.75$                             \\ 
        $\ExternalNode = 6$           & $0.75$ & $0.3$                              \\ 
    \end{tabular}
\end{table}

\subsection{DC switching station}
Up to four DCCBs are considered in the case study, which means that ${\NumberOfBreakers\! = 4}$ and ${\InternalNodeSet = \{1:5\}}$ in the optimization problem. The power flow through non-faulted DCCBs is limited by the constraints in~\eqref{eq:DCCB_power_normal} and~\eqref{eq:DCCB_power_cont}. However, detailed considerations on the steady-state power flow capacity through DCCBs are out of the scope of this paper. Therefore, these constraints are not considered in the case study.

Finally, the DCCB cost $\BreakerCost$ is an important parameter in the cost-benefit analysis. In the objective function, $\BreakerCost$ must present the total cost of a DCCB per time period considered in the optimization problem. Realistic values depend, therefore, on the total investment costs, the lifetime of the breaker, interest rates and operational costs. Since sufficient data for these values is not available in the public domain, the value of $\BreakerCost$ used in this paper is also defined based on the reference cost~$\ReferenceValue$, as shown in~\eqref{eq:Dccb_cost_reference}. This value is subsequently varied in a sensitivity analysis in Section~\ref{sec:sensitivity}.
\begin{equation}
    \label{eq:Dccb_cost_reference}
    \BreakerCost = 500\cdot\ReferenceValue
\end{equation}
\section{Case Study Results}
\label{sec:results}
This section presents the results of the breaker topology optimization method as applied to the case study described in the previous section. First, the risk-based cost-benefit analysis is performed to determine the optimal number of DCCBs and their configuration in the case study network shown in Fig.~\ref{fig:case_study_network}. Second, a sensitivity analysis is performed on the DCCB cost $\BreakerCost$ and the fault probability $\Probability_\Contingency$ to evaluate how these inputs affect the optimal placement and number of breakers in the DCSS. 

\subsection{Risk-based cost-benefit analysis}
\label{sec:CBA_case_study}
\begin{figure}[ht!]
    \centering
    \includegraphics[scale=0.85]{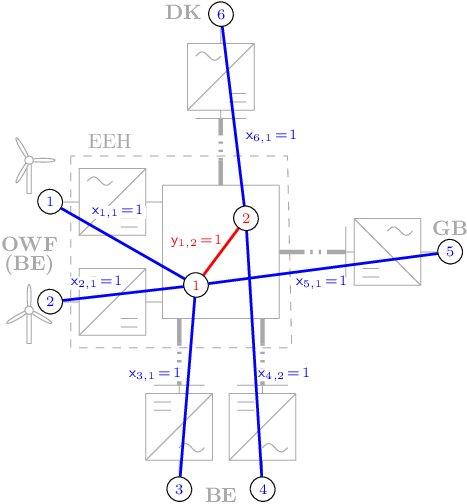}
    \caption{Optimal DCCB configuration for the considered DC switching station}
    \label{fig:cba_result1}
\end{figure}

Using the given values from Section~\ref{sec:case_study_introduction}, the topology resulting from the solution to the optimization problem is shown in Fig.~\ref{fig:cba_result1}. One DCCB is used, creating two DC protection zones. 

Although the connection to ${\ExternalNode\!=\!6}$ has the highest fault probability, due to the length of this connection, it is not fully separated in its own protection zone. Instead, another external node ${\ExternalNode\!=\!4}$ is connected to the same DC node, allowing uninterrupted power flow between these AC grids during faults on any of the connections to ${\InternalNode\!=\!1}$. In this optimal solution, the AC node with the highest imbalance costs, ${\ExternalNode\!=\!5}$, is connected to the DC node with the lowest total fault probability. 

For this result, the objective value amounts to $1386\:\ReferenceValue$. Given the assumed DCCB cost of $\BreakerCost=500\:\ReferenceValue$, this means that the breaker makes up $36.08\%$ of the total cost, while the other $63.92\%$ represents the fault impact risk.

\subsection{Sensitivity analysis}
\label{sec:sensitivity}
The solution of the risk-based cost-benefit analysis strongly depends on the applied cost values and the failure rates of the HVDC connections. Therefore, a sensitivity analysis is performed for these parameters. 

\begin{figure}[h]
    \centering
    \includegraphics[width=0.7\linewidth]{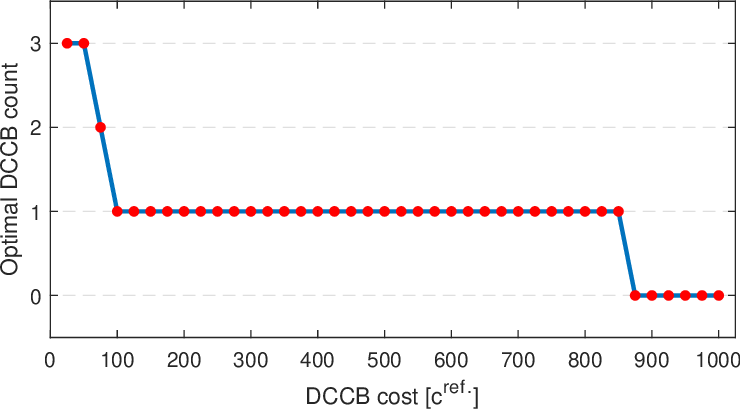}
    \caption{Optimal number of DCCBs based on the breaker cost}
    \label{fig:CB_cost_sensitivity}
\end{figure}

\begin{figure}[ht]
    \centering
    \begin{subfigure}{0.475\linewidth}
        \includegraphics[width=\linewidth]{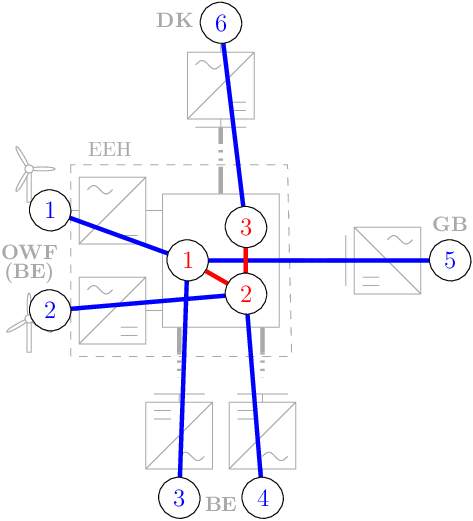} 
        \caption{}
        \label{fig:config_2cb}
    \end{subfigure}\hfill
    \begin{subfigure}{0.475\linewidth}
    \includegraphics[width=\linewidth]{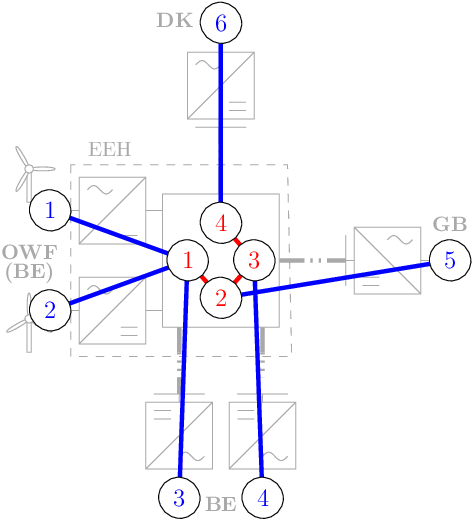}
    \caption{}
    \label{fig:config_3cb}
    \end{subfigure}\hfill
    \caption{Optimal DCCB configurations when 2 or 3 DCCBs are selected}
    \label{fig:optimal_config_multiple}
\end{figure}

First, the risk-based cost-benefit analysis is repeated for different values of $\BreakerCost$ (still expressed based on the reference cost $\ReferenceValue$). The result of this analysis, shown in Fig.~\ref{fig:CB_cost_sensitivity}, indicates that, if all other inputs and parameters are kept the same, the use of one DCCB in the configuration from Fig.~\ref{fig:cba_result1} is optimal for breaker costs between $100\:\ReferenceValue$ and $850\:\ReferenceValue$. For higher breaker costs, it becomes optimal to not include any DCCBs, accepting the total fault impact risk of $\riska\: \ReferenceValue$. Moreover, this sensitivity analysis shows that for DCCB costs lower than $100\:\ReferenceValue$, it becomes optimal to include a second or third DCCB in the DCSS. Fig.~\ref{fig:optimal_config_multiple} shows the optimal DCCB configurations for these breaker counts. 

The optimization method can be used to calculate the total risk for different numbers of optimally configured breakers. This total risk is represented by the second term of the objective function in~\eqref{eq:CBA_objective}. Based on these results, shown for up to four DCCBs in Tab.~\ref{tab:risk_values}, the \textit{marginal benefit} of each DCCB can be calculated based on how much each additional DCCB reduces the total fault-impact risk in the case study, when optimally configured. 

\begin{table}[ht!]
    \centering
    \caption{Fault-impact risk for different numbers of optimally configured DCCBs in the considered case study.}
    \label{tab:risk_values}
    \begin{tabular}{ c c c}
        DCCB count       & Total risk [$\ReferenceValue$] & DCCB marginal benefit [$
        \ReferenceValue$] \\ \hline
        0           & $\riska$& -                            \\
        1        & $\riskb$& $\cbone$                   \\ 
        2          & $\riskc$ & $\cbtwo$                        \\ 
        3        & $\riskd$ & $\cbthree$                              \\ 
        4          & $\riske$ & $\cbfour$                             \\ 
    \end{tabular}
\end{table}

\begin{figure}[ht!]
    \centering
    \includegraphics[width=0.7\linewidth]{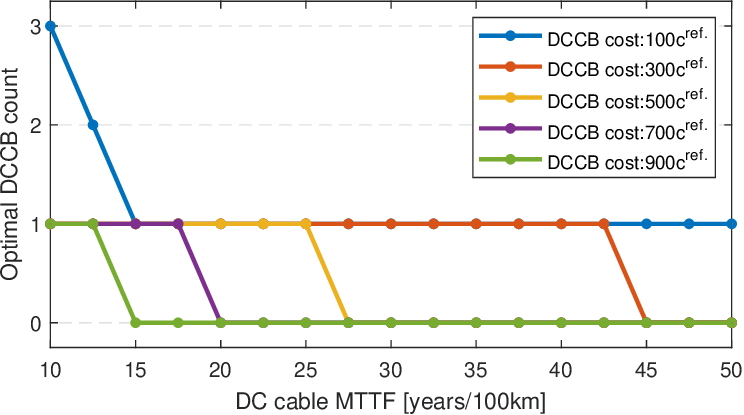}
    \caption{Optimal number of DCCBs based on DC cable MTTF for different breaker costs}
    \label{fig:reliability_sensitivity}
\end{figure}

The total fault-impact risk results listed in Tab.~\ref{tab:risk_values}, which determine the marginal benefit of DCCBs, are subject to the considered fault probabilities. These probabilities are calculated in~\eqref{eq:MTTF_to_prob} based on an example MTTF for HVDC cables. Therefore, since the risk calculated in the objective function is proportional to the assumed failure rate, higher MTTF values decrease the value of DCCBs for risk minimization in the DCSS. For example, if a MTTF of 50~years/$100\km$ is considered, as proposed in the \textit{best~case} scenario in~\cite{maciver2015reliability}, no DCCBs would be implemented in the result of the optimization problem (assuming a cost of $500\:\ReferenceValue$), since the total risk without any breakers amounts to $\risklowprobnocb$ and the use of one optimally configured DCCB reduces the risk only to $\risklowprobonecb$. Fig.~\ref{fig:reliability_sensitivity} shows the optimal DCCB count based on the considered failure rates for different breaker costs. 

\section{Conclusion}
\label{sec:conclusion}
The proposed method for HVDC circuit breaker (DCCB) configuration optimization in HVDC switching stations (DCSS) offers a novel approach to conceptual HVDC protection system design. The connectivity of external connections to the DCSS and the arrangement of DCCBs within the DCSS is optimized to minimize  protection system costs and the total fault-impact risk. Consequently, this method ensures that future HVDC grids can be developed with cost-effective protection systems, optimized for case-specific operating conditions. This improves on the current practice that relies on general, pre-defined protection strategies. Moreover, the consideration of fault-impact risk in the objective function allows protection systems to be optimized beyond current maximum loss-of-infeed limits, ensuring the effectiveness of these systems in the future, when the reduction of conventional generation capacity might cause these limits to change.

The optimization method was applied to a case study with six external connections using pre-fault power flow scenarios representing one year of operation, calculated using economic dispatch optimization. While the specific outcomes discussed in this paper are case-study specific, it can be concluded that the proposed method is generally applicable for calculating the optimal number of DCCBs in a DCSS, while also determining the optimal configuration thereof. Additionally, the conducted sensitivity analyses of the DCCB costs and DC cable failure rates provide a clear view of how these parameters influence the optimal DCCB count and the total fault-impact risk. Consequently, the proposed method can serve not only as a tool for optimal DCSS design, but is also useful for studying the benefits of selective DC protection systems conceptually. 

In this paper, the optimization problem considers a single, centralized DCSS. Future extensions of the formulation could include the co-design of multiple switching stations to represent meshed multiterminal HVDC grids. Moreover, for the industrial application of this method, it is possible to include detailed, time-dependent representations of imbalance costs and fault probability.

\section*{Acknowledgement}
This work is part of the DIRECTIONS project, supported by the Energy Transition Fund, FOD Economy, Belgium and partially supported by the project `Innovative solutions for underground high-voltage lines and grids', funded by the Flemish Government.

\bibliography{References.bib}

\end{document}